\documentclass[preprint,aps,  nofootinbib]{revtex4}
\usepackage{amssymb}
\usepackage{amsmath}
\usepackage{graphicx}

\newcommand\beq{ \begin{eqnarray} }
\newcommand\eeq{ \end{eqnarray} }

\begin{document}

\title{Bulk Viscosity of a Gas of Massless Pions}
\author{Jiunn-Wei Chen$^{1,2}$ and Juven Wang$^{1}$}
\affiliation{$^{1}$ Department of Physics and Center for Theoretical Sciences, National
Taiwan University, Taipei 10617}
\affiliation{$^{2}$ CTP, Massachusetts Institute of Technology, Cambridge, MA 02139}

\begin{abstract}
In the hadronic phase, the dominant configuration of QCD with two flavors of
massless quarks is a gas of massless pions. We calculate the bulk viscosity (%
$\zeta $) using the Boltzmann equation with the kinetic theory generalized
to incorporate the trace anomaly. We find that the dimensionless ratio $%
\zeta /s$, $s$ being the entropy density, is monotonic increasing below $T=$%
120 MeV, where chiral perturbation theory is applicable. This, combined with
previous results, shows that $\zeta /s$ reaches its maximum near the phase
transition temperature $T_{c}$, while $\eta /s$, $\eta $ being the shear
viscosity, reaches its minimum near $T_{c}$ in QCD with massless quarks.
\end{abstract}

\maketitle


\section{Introduction}

Transport coefficients of Quantum Chromodynamics (QCD) are of high interests
recently. This was triggered by the discovery that quark gluon plasma (QGP)
has a viscosity close to the conjectured universal minimum bound \cite{KOVT1}%
, indicating that QGP is close to a \textquotedblleft perfect
fluid\textquotedblright\ \cite{RHIC,Molnar:2001ux,Teaney:2003pb}\ just above
the deconfinement temperature. This bound, $\eta /s$ $\geq 1/4\pi $, $s$
being the entropy density, is motivated by the uncertainty principle and is
found to be saturated for a large class of strongly interacting quantum
field theories whose dual descriptions in string theory involve black holes
in anti-de Sitter space \cite%
{Policastro:2001yc,Policastro:2002se,Herzog:2002fn,Buchel:2003tz}. There are
some debates about whether the minimum bound on $\eta /s$ is truly universal 
\cite{Cohen:2007qr,Dobado:2007tm,Son:2007xw} and the RHIC data might be
better fit with $\eta /s$ $<1/4\pi $ \cite{Romatschke:2007mq,Song:2007fn}
(lattice results for gluon plasma, however, is still consistent with the
bound \cite{etas-gluon-lat,etas-gluon-lat-x}). In any case, smaller $\eta $
implies stronger interparticle interaction (here $\eta $ is normalized by
the density) and the smallness of QGP $\eta $ indicating an intriguing
strongly interacting state is reached near the deconfinement temperature.

In general, the minimum of $\eta /s$ is found near the phase transition
temperature $T_{c}$ or when the system goes through a fast crossover. This
behavior was observed \cite{KOVT1,Csernai:2006zz,Chen:2007xe}\ in all the
materials, including N, He, and H$_{2}$O, with data available in the NIST
and CODATA websites \cite{webbook,codata}. Surprisingly, it is also observed
in QCD at zero chemical potential \cite{Csernai:2006zz,Chen:2006iga}, near
the nuclear liquid gas phase transition \cite{Chen:2007xe,Itakura:2007mx},
and in cold fermionic atom systems at the limit with two-body scattering
length tuned to infinity \cite{etas-supfluid}. Using weakly-coupled real
scalar field theories, in which perturbation is reliable, the same $\eta /s$
behaviors in first-, second-order phase transitions and crossover also
emerge as in the liquid-gas transitions in N, He, and H$_{2}$O and
essentially all the matters with data available in the NIST database
mentioned above \cite{Chen:2007jq}. This agreement is expected to hold when
the theory is generalized to $N$ components with an $O(N)$ symmetry. Thus,
these behaviors might be general properties of fluid and might be used to
probe the QCD critical end point \cite{Lacey:2006bc}.

Less well studied is the bulk viscosity ($\zeta $) of QCD. In general, bulk
viscosity vanishes when a system is conformally invariant such that the
system is invariant under a uniform expansion (dilatation). For a
non-interacting non-relativistic or ultrarelativistic system (assuming the
interaction is turned off after thermal equilibrium), the system is
conformally invariant and hence has zero bulk viscosity. When the
interaction is turned on, conformal symmetry could be broken to give a
finite bulk viscosity. (A notable exception is the infinite scattering
length limit where conformal symmetry is preserved \cite%
{Mehen:1999nd,Son:2005rv,Son:2005tj}.) In QCD with heavy quarks integrated
out and with the light quark masses set to zero, conformal symmetry is
broken in the quantum level. In the perturbative region, up to some
logarithmic corrections, $\zeta /s\propto \alpha _{s}^{-2}\left(
1/3-v_{s}^{2}\right) ^{2}\propto \alpha _{s}^{2}$ \cite{Arnold:2006fz}\
while $\eta /s\propto \alpha _{s}^{-2}$ \cite{Arnold:2000dr,Arnold:2003zc}.
Thus, $\zeta $ is smaller than $\eta $ in the perturbative regime. When the
temperature is reduced, $\eta /s$ reaches its minimum near $T_{c}$, while $%
\zeta /s$ rises sharply near $T_{c}$ \cite%
{Kharzeev:2007wb,Karsch:2007jc,Meyer:2007dy}. It will be interesting to see
whether the maximum of $\zeta /s$ is also reached near $T_{c}$ from below,
which is the main purpose of this work. We will focus on the case with two
flavors of massless quarks such that below $T_{c}$ the dominant degrees of
freedom are massless pions.

\section{Linearized Boltzmann Equation and the Generalized Kinematic Theory}

The bulk viscosity of a system is defined by the Kubo formula%
\begin{equation}
\zeta =\frac{1}{9}\lim_{\omega \rightarrow 0}\frac{1}{\omega }%
\int_{0}^{\infty }dt\int d^{3}r\,e^{i\omega t}\,\langle \lbrack T_{\mu
}^{\mu }(x),T_{\nu }^{\nu }(0)]\rangle \,,
\end{equation}%
with $T_{\mu }^{\mu }$ the trace of the energy momentum tensor. The Kubo
formula involves an infinite number of diagrams at the leading order (LO)
even in the weakly-coupled $\phi ^{4}$ theory \cite{Jeon}. However, it is
proven that the summation of LO diagrams in a weakly coupled $\phi ^{4}$
theory \cite{Jeon} or in hot QED \cite{Gagnon:2007qt,Gagnon:2006hi} is
equivalent to solving the linearized Boltzmann equation with
temperature-dependent particle masses and scattering amplitudes. Since the
proofs do not use properties restricted to scalar theories, the conclusion
is expected to hold for more general theories with weak couplings, including
QCD in the perturbative regime \cite%
{Arnold:2006fz,Arnold:2000dr,Arnold:2003zc}. Here, we assume the equivalence
between the Kubo formula and the Boltzmann equation also applies to massless
pions.

The Boltzmann equation describes the evolution of the isospin averaged pion
distribution function $f=f(\mathbf{x},\mathbf{p},t)\equiv f_{p}(x)$ (a
function of space, time and momentum) 
\begin{equation}
\frac{p^{\mu }}{E_{p}}\partial _{\mu }f_{p}(x)=\frac{g_{\pi }}{2}%
\int_{123}d\Gamma _{12;3p}\left\{
f_{1}f_{2}(1+f_{3})(1+f_{p})-(1+f_{1})(1+f_{2})f_{3}f_{p}\right\} \ ,
\label{Bz}
\end{equation}%
where $E_{p}=\sqrt{p^{2}+m_{\pi }^{2}}$, $p=\left\vert \mathbf{p}\right\vert 
$ and $g_{\pi }=3$ is the degeneracy factor for three pions , 
\begin{equation}
d\Gamma _{12;3p}\equiv \frac{1}{2E_{p}}|\mathcal{T}|^{2}\prod_{i=1}^{3}\frac{%
d^{3}\mathbf{k}_{i}}{(2\pi )^{3}(2E_{i})}\times (2\pi )^{4}\delta
^{4}(k_{1}+k_{2}-k_{3}-p)\ ,  \label{Bz1}
\end{equation}%
and where $\mathcal{T}$ is the scattering amplitude for particles with
momenta $1,2\rightarrow 3,p$. In chiral perturbation theory ($\chi $PT),
which is a low-energy effective field theory of QCD, the LO isospin averaged 
$\pi \pi $ scattering amplitude in terms of Mandelstam variables ($s,t$, and 
$u$) is 
\begin{equation}
|\mathcal{T}|^{2}=\frac{1}{9f_{\pi }^{4}}\left\{ 9s^{2}+3(t-u)^{2}\right\} \
,
\end{equation}%
where $f_{\pi }=88.3$ MeV is the pion decay constant in the chiral limit.
The pions remain massless below $T_{c}$ and the temperature dependence of
the scattering amplitude is of higher order and will be neglected.

In local thermal equilibrium, the distribution function $f_{p}^{(0)}(x)=%
\left( e^{\beta (x)V_{\mu }(x)p^{\mu }}-1\right) ^{-1}$, where $\beta
(x)=1/T(x)$ is the inverse temperature and $V^{\mu }(x)$ is the four
velocity of the fluid at the space-time point $x$. A small deviation of $%
f_{p}$ from local equilibrium is parametrized as 
\begin{eqnarray}
f_{p}(x) &=&f_{p}^{(0)}(x)+\delta f_{p}(x)\ ,  \notag \\
\delta f_{p}(x) &=&-f_{p}^{(0)}(x)\left[ 1+f_{p}^{(0)}(x)\right] \chi
_{p}(x)\ .
\end{eqnarray}

In kinetic theory, the energy momentum tensor in a weakly interacting system
is 
\begin{equation}
T_{\mu \nu }(x)=g_{\pi }\int \frac{\mathrm{d}^{3}\mathbf{p}}{(2\pi )^{3}}%
\frac{f_{p}(x)}{E_{p}}p_{\mu }p_{\nu }\ .  \label{2}
\end{equation}%
It is the sum of the energy momentum tensor of each particle with
inter-particle interactions neglected. This is usually a good approximation
when the interparticle spacing is much larger than the range of interaction
such that the potential energy is negligible.

The conservation of energy momentum tensor, $\partial ^{\mu }T_{\mu \nu }=0$%
, is automatically satisfied by the Boltzmann equation. We will decompose $%
T_{\mu \nu }$ as 
\begin{equation}
T_{\mu \nu }=T_{\mu \nu }^{(0)}+\delta T_{\mu \nu }\ ,
\end{equation}%
where $\delta T_{\mu \nu }$ is the deviation from the thermal equilibrium
part $T_{\mu \nu }^{(0)}$. 
\begin{equation}
T_{\mu \nu }^{(0)}=\left( \epsilon +{\mathcal{P}}\right) V_{\mu }V_{\nu }-{%
\mathcal{P}}g_{\mu \nu }\ ,
\end{equation}%
where $\epsilon $ is the energy density and ${\mathcal{P}}$ is the pressure.

We will work at the $\mathbf{V}(x)=0$ frame for the point $x$. This implies $%
\partial _{\nu }V^{0}=0$ after taking a derivative on $V_{\mu }(x)V^{\mu
}(x)=1$. The conservation law at local thermal equilibrium, $\partial ^{\mu
}T_{\mu \nu }^{(0)}=0$, implies%
\begin{eqnarray}
\partial _{t}\epsilon +\left( \epsilon +{\mathcal{P}}\right) \nabla \cdot 
\mathbf{V} &=&0\ ,  \notag \\
\partial _{t}\mathbf{V}+\left( \epsilon +{\mathcal{P}}\right) ^{-1}\nabla {%
\mathcal{P}} &=&0\ .  \label{Shu-Fen}
\end{eqnarray}%
Then using the thermal dynamic relation%
\begin{equation}
\epsilon +{\mathcal{P}}=T\frac{\partial {\mathcal{P}}}{\partial T}\ ,
\end{equation}%
one has%
\begin{eqnarray}
\beta \partial _{t}\mathbf{V}-\mathbf{\nabla }\beta &=&0\ ,  \notag \\
\partial _{t}\beta -\beta v_{s}^{2}\nabla \cdot \mathbf{V} &=&0\ ,
\label{Yo-Lei}
\end{eqnarray}%
where $v_{s}^{2}=\partial {\mathcal{P}}/\partial \epsilon $ is the speed of
sound.\ 

The shear and bulk viscosity are defined by the small deviation away from
equilibrium: 
\begin{equation}
\delta T_{ij}=-2\eta \left( \frac{\nabla _{i}V_{j}(x)+\nabla _{j}V_{i}(x)}{2}%
-\frac{1}{3}\delta _{ij}\nabla \cdot \mathbf{V}(x)\right) -\zeta \delta
_{ij}\nabla \cdot \mathbf{V}(x)\ ,  \label{Yo-Hao}
\end{equation}%
where $i$ and $j$ are spacial indexes and Eq.(\ref{Yo-Lei}) is used to
replace the time derivatives $\partial _{t}\beta $ and $\partial _{t}\mathbf{%
V}$ by spacial derivatives $\nabla \cdot \mathbf{V}$ and $\mathbf{\nabla }%
\beta $. Also, $\delta T_{0i}(x)=0$, since the momentum density at point $x$
is zero in the $\mathbf{V}(x)=0$ frame. Furthermore, if there is no
viscosity, the energy density at the same point will only be a function of $%
T $ governed by thermodynamics, which implies $\delta T_{00}=0$. Viscosity
could generate heat during the perturbation. However, the amount of heat
generated should be time reversal even, because heat will be generated no
matter whether the system is expanding or contracting. However, there is no
first derivative term which is even under time reversal. Thus, at this
order, 
\begin{equation}
\delta T_{00}=0\ .  \label{T00x}
\end{equation}

It is easy to see why $\zeta \simeq 0$ for ultrarelativistic and monatomic
non-relativistic systems based on Eqs. (\ref{2}) and (\ref{T00x}). For
ultrarelativistic systems, $p^{2}\simeq 0$; therefore, $T_{\mu }^{\mu
}\simeq 0$ by Eq. (\ref{2}). For non-relativistic systems, if the particle
number for each species is conserved, then $\delta T_{i}^{i}=2\delta
T_{0}^{0}=0$ and, hence, $\zeta =0$. These are general results of the
kinetic theory which assumes the potential energy from short-range
interactions is negligible in a dilute system. They can be traced back to
the conformal symmetry of non-interacting ultrarelativistic and
non-relativistic systems. When interactions are turned on and the conformal
symmetry is broken, Eq.(\ref{2}) has to be modified to include the effect of
interaction in order to give the leading non-vanishing $\zeta $ result.

For pions in the chiral limit, they always satisfy the dispersion relation $%
p^{2}=0$ even at finite T. This is because their goldstone boson nature
prevents them from generating thermal masses. However, this does not imply
that the system is traceless. Direct computation using $\chi $PT shows that
trace anomaly first appears at the order of three loops \cite{Gerber:1988tt}%
. This is the manifestation of the gluon trace anomaly operator of QCD. In
the expression of Eq.(\ref{2}), $T_{\mu }^{\mu }=0$ once $p^{2}=0$. Thus, it
needs to be generalized to have non-zero $T_{\mu }^{\mu }$. In principle,
one could add two-pion, three-pion... distribution amplitudes to take into
account the pion interaction associate with the loop diagrams. However, one
can integrate out the medium effect and sum up the effective one-pion
contributions to $T_{\mu \nu }$ 
\begin{equation}
T_{\mu \nu }=\sum_{i}\left\langle \pi _{i}\left\vert \widehat{T}_{\mu \nu
}\right\vert \pi _{i}\right\rangle \ ,  \label{MIT}
\end{equation}%
where $\widehat{T}_{\mu \nu }$ is the energy momentum operator. Note that
Eq. (\ref{2}) is just the leading order effect of the above equation which
takes into account the free pion contribution to $T_{\mu \nu }$ only. Using
symmetries, $T_{\mu \nu }$ has the general form:%
\begin{equation}
T_{\mu \nu }(x)=g_{\pi }\int \frac{\mathrm{d}^{3}\mathbf{p}}{(2\pi )^{3}}%
\frac{f_{p}(x)}{E_{p}}\left[ p_{\mu }p_{\nu }\left( 1+g_{1}(x)\right) +\frac{%
g_{2}(x)g_{\mu \nu }}{\beta (x)^{2}}+\frac{g_{3}(x)V_{\mu }(x)V_{\nu }(x)}{%
\beta (x)^{2}}\right] \ .  \label{T_mu_nu}
\end{equation}%
Here Lorentz symmetry is broken down to $O(3)$ symmetry by the temperature,
and $g_{1-3}$ are dimensionless functions of $\beta (x)$ and $f_{\pi }$. In $%
\chi $PT, $g_{1-3}=\mathcal{O}(T^{4}/(4\pi f_{\pi })^{4})$ \cite%
{Gerber:1988tt}. The structure $\left( p^{\mu }V^{\nu }+V^{\mu }p^{\nu
}\right) $ is not allowed because the $\pi ^{+}$ and $\pi ^{-}$ matrix
elements should be the same by charge conjugation or isospin symmetry. Thus, 
$\left\langle \pi _{i}(p)\left\vert \widehat{T}_{\mu \nu }\right\vert \pi
_{i}(p)\right\rangle $ should be invariant under crossing symmetry ( $p^{\mu
}\rightarrow -p^{\mu }$). In equilibrium, $T_{\mu }^{(0)\mu }=\epsilon -3{%
\mathcal{P}}$ and%
\begin{equation}
c\equiv 4g_{2}+g_{3}=\frac{\epsilon -3{\mathcal{P}}}{\dfrac{g_{\pi }}{\beta
^{2}}\int \dfrac{\mathrm{d}^{3}\mathbf{p}}{(2\pi )^{3}}\dfrac{f_{p}^{(0)}}{%
E_{p}}}\ .  \label{C}
\end{equation}

Note that energy momentum conservation is not a problem with the new terms
in Eq.(\ref{T_mu_nu}). In equilibrium, one just has to replace $v_{s}^{2}$
in Eq.(\ref{Yo-Lei}) by the new value to obtain $\partial ^{\mu }T_{\mu \nu
}^{(0)}=0$. Away from equilibrium, the net effect of $\zeta $ is to replace $%
{\mathcal{P}}\rightarrow {\mathcal{P}}-\zeta \nabla \cdot \mathbf{V}$ in Eq.(%
\ref{Shu-Fen}) which will induce second spacial derivative terms in Eq.(\ref%
{Yo-Lei}). Thus, as long as Eq.(\ref{T_mu_nu}) gives the correct $T_{\mu \nu
}$, energy momentum conservation can be satisfied.

\bigskip Working to the first order in a derivative expansion, $\chi _{p}(x)$
can be parametrized as 
\begin{equation}
\chi _{p}(x)=\beta (x)A(p)\nabla \cdot \mathbf{V}(x)+\beta (x)B(p)\left( 
\hat{p}_{i}\hat{p}_{j}-\frac{1}{3}\delta _{ij}\right) \left( \frac{\nabla
_{i}V_{j}(x)+\nabla _{j}V_{i}(x)}{2}-\frac{1}{3}\delta _{ij}\nabla \cdot 
\mathbf{V}(x)\right) \ ,  \label{df1}
\end{equation}%
where $A$ and $B$ are functions of $x$ and $p$. But we have suppressed the $%
x $ dependence. Substituting (\ref{df1}) into the Boltzmann equation and
using Eq. (\ref{Yo-Lei}), one obtains one linearized equation for $A$
(associated with the $\nabla \cdot \mathbf{V}$ structure): 
\begin{eqnarray}
\frac{1}{3}p^{2}-v_{s}^{2}E_{p}^{2} &=&\frac{g_{\pi }E_{p}}{2}%
\int_{123}d\Gamma _{12;3p}(1+n_{1})(1+n_{2})n_{3}(1+n_{p})^{-1}  \notag \\
&&\times \left[ A(p)+A(k_{3})-A(k_{2})-A(k_{1})\right] \ ,  \label{A}
\end{eqnarray}%
where at point $x$, $f_{i}^{(0)}(x)$ is written as $n_{i}=\left( e^{\beta
E_{i}}-1\right) ^{-1}$. There is also a linearized equation for $B$
(associated with the $\left( \nabla _{i}V_{j}+\nabla _{j}V_{i}-\text{trace}%
\right) $ structure) that is related to the shear viscosity $\eta $. The
computation of $\eta $ of the pion gas has been discussed in Ref. \cite%
{Chen:2006iga}. We will focus on solving $\zeta $ in this work.

\section{Variational Calculation}

Equation (\ref{A}) only determines $A(p)$ up to a combination $%
a_{1}+a_{2}E_{p}$, where $a_{1}$ and $a_{2}$ are constants \cite{Jeon}.
These \textquotedblleft zero modes\textquotedblright\ ($a_{1}$ and $%
a_{2}E_{p}$) only appear in the analysis of bulk viscosity but not shear
viscosity. We will discuss their effects in this section.

The variation of Eq. (\ref{T_mu_nu}) yields

\begin{eqnarray}
\delta T_{\mu \nu } &=&g_{\pi }\int \frac{\mathrm{d}^{3}\mathbf{p}}{(2\pi
)^{3}E_{p}}\left\{ \delta f_{p}\left[ p_{\mu }p_{\nu }\left( 1+g_{1}\right) +%
\frac{g_{2}g_{\mu \nu }}{\beta ^{2}}+\frac{g_{3}V_{\mu }V_{\nu }}{\beta ^{2}}%
\right] \right.   \notag \\
&&+\left. f_{p}\left[ p_{\mu }p_{\nu }\delta g_{1}+\frac{\delta g_{2}g_{\mu
\nu }}{\beta ^{2}}+\frac{\delta g_{3}V_{\mu }V_{\nu }}{\beta ^{2}}\right]
\right\} \ .
\end{eqnarray}%
Note that $g_{1-3}$ represent loop corrections of the energy momentum
tensor, thus they are functionals of $f_{p}$. To compute $\zeta $, we need  
\begin{equation}
\delta T_{ii}=g_{\pi }\int \frac{\mathrm{d}^{3}\mathbf{p}}{(2\pi )^{3}E_{p}}%
\left\{ \delta f_{p}\left[ p^{2}\left( 1+g_{1}\right) -\frac{3g_{2}}{\beta
^{2}}\right] +f_{p}\left[ p^{2}\delta g_{1}-\frac{3\delta g_{2}}{\beta ^{2}}%
\right] \right\} \ .  \label{G}
\end{equation}%
This can be simplified using the constraint,  \  
\begin{equation}
0=\delta T_{00}=g_{\pi }\int \frac{\mathrm{d}^{3}\mathbf{p}}{(2\pi )^{3}E_{p}%
}\left\{ \delta f_{p}\left[ p^{2}\left( 1+g_{1}\right) +\frac{g_{2}+g_{3}}{%
\beta ^{2}}\right] +f_{p}\left[ p^{2}\delta g_{1}+\frac{\delta g_{2}+\delta
g_{3}}{\beta ^{2}}\right] \right\} \ .  \label{F}
\end{equation}%
After eliminating the $g_{2}/\beta ^{2}$ term in $\delta T_{ii}$ using the
constraint, we have%
\begin{eqnarray}
\delta T_{ii} &=&g_{\pi }\int \frac{\mathrm{d}^{3}\mathbf{p}}{(2\pi
)^{3}E_{p}}\left\{ \delta f_{p}\left[ 4p^{2}\left( 1+g_{1}\right) \frac{%
4g_{2}+g_{3}}{g_{2}+g_{3}}\right] \right.   \notag \\
&&+\left. f_{p}\left[ p^{2}\delta g_{1}\frac{4g_{2}+g_{3}}{g_{2}+g_{3}}+%
\frac{3\left( g_{2}\delta g_{3}-g_{3}\delta g_{2}\right) }{\left(
g_{2}+g_{3}\right) \beta ^{2}}\right] \right\} \   \notag \\
&\simeq &4g_{\pi }d\int \frac{\mathrm{d}^{3}\mathbf{p}}{(2\pi )^{3}}p\delta
f_{p}\left( 1+\mathcal{O}(\frac{T^{4}}{(4\pi f_{\pi })^{4}})\right) \ ,
\end{eqnarray}%
where $d=\left( 4g_{2}+g_{3}\right) /\left( g_{2}+g_{3}\right) $ and the
pion remains massless in the chiral limit even at finite $T$, so we have
used $p^{2}=E_{p}^{2}$. The above expression for $\delta T_{ii}$ implies 
\begin{equation}
\zeta =\frac{4}{3}g_{\pi }\beta d\int \frac{\mathrm{d}^{3}\mathbf{p}}{(2\pi
)^{3}}E_{p}n_{p}\left( 1+n_{p}\right) A(p)\ .  \label{D}
\end{equation}%
Then using Eq. (\ref{A}) and the symmetry property of the scattering
amplitude, 
\begin{eqnarray}
\zeta  &=&\frac{g_{\pi }^{2}\beta d}{2\left( 1-3v_{s}^{2}\right) }\int
\prod_{i=1,2,3,p}\frac{d^{3}\mathbf{k}_{i}}{(2\pi )^{3}(2E_{i})}|\mathcal{T}%
|^{2}(2\pi )^{4}\delta ^{4}(k_{1}+k_{2}-k_{3}-p)  \notag \\
&&\times (1+n_{1})(1+n_{2})n_{3}n_{p}\left[ A(p)+A(k_{3})-A(k_{2})-A(k_{1})%
\right] ^{2}\ .  \label{E}
\end{eqnarray}

Note that equating Eqs. (\ref{D}) and (\ref{E}) is equivalent to taking a
projection of Eq. (\ref{A}). It can be shown that any ansatz satisfying Eqs.
(\ref{D}) and (\ref{E}) gives a lower bound on $\zeta $ \cite{Resibois}.
Thus, one can solve $\zeta $ variationally, i.e. finding an ansatz $A(p)$
that gives the biggest $\zeta $.

It is known that if one uses the ansatz $A(p)=a_{1}+a_{2}E_{p}$, then it
will not contribute to the $2\rightarrow 2$ scattering on the right-hand
side of Eq. (\ref{A}) (the $a_{2}$ terms cancel by energy conservation). In
fact, this ansatz will not contribute to all the particle number conserving
processes but can contribute to particle number changing processes, such as $%
2\leftrightarrow 4$ scattering, which we have not shown. As we know from
Eqs. (\ref{A}) and (\ref{D}), $\zeta $ is proportional to the size of $A(p)$
which is inversely proportional to rate of scattering. \bigskip Thus, if the 
$2\rightarrow 2$ scattering has a bigger rate than the $2\leftrightarrow 4$
scattering, then this ansatz gives a bigger $\zeta $ by bypassing the faster 
$2\rightarrow 2$ scattering. In $\phi ^{4}$ theory, it was found that $\zeta 
$ is indeed set by the $2\leftrightarrow 4$ scattering \cite{Jeon}. However,
in perturbative QCD (PQCD), the soft particle number changing bremsstrahlung
is faster than the $2\rightarrow 2$ scattering \cite{Arnold:2006fz}. Thus, $%
\zeta $ is governed by $2\rightarrow 2$ scattering.

In the case with massless pions, however, $2\rightarrow 2$ scattering is
still the dominant process. While using the ansatz $A(p)=a_{1}+a_{2}E_{p}$,
the $\delta T_{00}=0$ constraint in Eq.(\ref{F}) demands $a_{1}/a_{2}=0$
because $n_{p}\propto 1/p$ as $p\rightarrow 0$. Since $A(p)$ parametrizes a
small deviation of $f_{p}$ away from thermal equilibrium, $a_{1}/a_{2}=0$
gives $a_{1}=0$ instead of $a_{2}\rightarrow \infty $ and $a_{1}$ finite.
Thus, to maximize $\zeta $, we uses the ansatz $%
A(p)=a_{2}E_{p}+a_{3}E_{p}^{2}+...$ without the $a_{1}$ term. The point is, $%
2\rightarrow 2$ scattering cannot be bypassed and it will be the dominant
process in our calculation.

\bigskip To compute $\zeta $, it is easier to eliminate the $\left(
1+g_{1}\right) $ term in Eq.(\ref{G}) using Eq.(\ref{F}):

\begin{eqnarray}
\delta T_{ii} &=&-g_{\pi }\int \frac{\mathrm{d}^{3}\mathbf{p}}{(2\pi
)^{3}E_{p}}\left\{ \delta f_{p}\left[ \frac{4g_{2}+g_{3}}{\beta ^{2}}\right]
\right.   \notag \\
&&+\left. f_{p}\left[ \frac{4\delta g_{2}+\delta g_{3}}{\beta ^{2}}\right]
\right\} \ .
\end{eqnarray}%
Note that $g_{2}$ and $g_{3}$ terms at $\mathcal{O}(T^{4}/(4\pi f_{\pi
})^{4})$ arise from three-loop diagrams and from two-loop diagrams with
insertions of higher order counterterms and each loop integral has one power
of $f_{p}$ in the integrand. Thus, we will make an approximation here to
assume the $\left( 4\delta g_{2}+\delta g_{3}\right) $ term is proportional
to the $\delta f_{p}$ term with a proportional constant $\left( l-1\right) $%
, where $l$ means the power of $f_{p}$ (or the number of loops) in $T_{ii}$.
Since $l$ is between $2$ and $3$, we take the mean value $l=2.5$ and
associate the uncertainty of $l$ to the error estimation of $\zeta $. 

The trace anomaly for massless pions appears from three-loop diagrams and
from two-loop diagrams with insertions of higher order counterterms [all are 
$\mathcal{O}(T^{8}/f_{\pi }^{4})$] \cite{Gerber:1988tt}. Thus,

For two-loop diagrams, the associated $l$ factor is $2$ while for three-loop
diagrams, the associated $l$ factor is $3$. Here, without distinguishing the
contribution from each diagram, we take the mean value $l=2.5$ and associate
the uncertainty of $l$ to the error estimation of $\zeta $. Thus, 
\begin{equation}
\zeta =-\frac{g_{\pi }lc}{3\beta }\int \frac{\mathrm{d}^{3}\mathbf{p}}{(2\pi
)^{3}}\frac{1}{E_{p}}n_{p}\left( 1+n_{p}\right) A(p)\ .
\end{equation}%
Note that $A(p)\propto g_{\pi }^{-1}\left( \frac{1}{3}-v_{s}^{2}\right)
f_{\pi }^{4}$ from Eq. (\ref{A}). Thus, for massless pions, 
\begin{equation}
\zeta =hl\left( \epsilon -3{\mathcal{P}}\right) \left( \frac{1}{3}%
-v_{s}^{2}\right) \frac{f_{\pi }^{4}}{T^{5}}\ ,
\end{equation}%
where $T^{5}$ is given by dimensional analysis and $h$ is a dimensionless
constant. To find the numerical solution for $h$, we neglect the
higher-order $g_{1-3}$ terms in Eq.(\ref{F}) and use the ansatz $%
A(p)=\sum_{n=1}^{m}c_{n}p^{n}$. We find 
\begin{equation}
h\simeq 65\ .
\end{equation}%
Using the $\chi $PT result of Ref. \cite{Gerber:1988tt} for $\epsilon $ and $%
{\mathcal{P}}$, we obtain

\begin{equation}
\zeta \simeq 0.15\left( \frac{l}{2.5}\right) \left( \ln \frac{\Lambda _{p}}{T%
}-\frac{1}{4}\right) \left( \ln \frac{\Lambda _{p}}{T}-\frac{3}{8}\right) 
\frac{T^{7}}{f_{\pi }^{4}}\ ,
\end{equation}%
where $\Lambda _{p}\simeq 275$ MeV. As expected, the bulk viscosity vanishes
as $f_{\pi }\rightarrow \infty $ or when the coupling between pions vanishes.

The leading order contribution for pion entropy density $s$ is just the
result for a free pion gas:%
\begin{equation}
s=\frac{2\pi ^{2}g_{\pi }}{45}T^{3}\ .
\end{equation}

\bigskip 

The trace anomaly for massless pions appears from three-loop diagrams and
from two-loop diagrams with insertions of higher order counterterms [all are 
$\mathcal{O}(T^{8}/f_{\pi }^{4})$] \cite{Gerber:1988tt}. Thus,

For two-loop diagrams, the associated $l$ factor is $2$ while for three-loop
diagrams, the associated $l$ factor is $3$. Here, without distinguishing the
contribution from each diagram, we take the mean value $l=2.5$ and associate
the uncertainty of $l$ to the error estimation of $\zeta $.%
\begin{equation*}
\delta T_{\mu \nu }\simeq g_{\pi }l\int \frac{\mathrm{d}^{3}\mathbf{p}}{%
(2\pi )^{3}E_{p}}\delta f_{p}\left[ p_{\mu }p_{\nu }\left( 1+g_{1}\right) +%
\frac{g_{2}g_{\mu \nu }}{\beta ^{2}}+\frac{g_{3}V_{\mu }V_{\nu }}{\beta ^{2}}%
\right] .
\end{equation*}%
Using Eqs. (\ref{T_mu_nu}) and (\ref{Yo-Hao}), 
\begin{eqnarray}
\delta T_{ii}(x) &=&-g_{\pi }\int \frac{\mathrm{d}^{3}\mathbf{p}}{(2\pi
)^{3}E_{p}}\left\{ \delta f_{p}\left[ \frac{4g_{2}+g_{3}}{\beta (x)^{2}}%
\right] \right.  \\
&&+\left. f_{p}\left[ \frac{4\delta g_{2}+\delta g_{3}}{\beta (x)^{2}}\right]
\right\} \ .
\end{eqnarray}%
\begin{eqnarray}
\delta T_{ii}(x) &=&g_{\pi }\int \frac{\mathrm{d}^{3}\mathbf{p}}{(2\pi
)^{3}E_{p}}\left\{ \delta f_{p}\left[ 4p^{2}\left( 1+g_{1}\right) \left(
4g_{2}+g_{3}\right) \right] \right.  \\
&&+\left. f_{p}(x)\left[ \left( 4g_{2}+g_{3}\right) p^{2}\delta g_{1}+\frac{%
3\left( g_{2}\delta g_{3}-g_{3}\delta g_{2}\right) }{\beta ^{2}}\right]
\right\} \ .
\end{eqnarray}%
where the factor $l$ appears because in the computation of $\delta T_{ii}$
from Eq. (\ref{T_mu_nu}), not only $f_{p}$ but also $g_{1-3}$ depend on $%
\delta f_{p}$. In $\chi $PT, the trace anomaly for massless pions appears
from three-loop diagrams and from two-loop diagrams with insertions of
higher order counterterms [all are $\mathcal{O}(T^{8}/f_{\pi }^{4})$] \cite%
{Gerber:1988tt}. For two-loop diagrams, the associated $l$ factor is $2$
while for three-loop diagrams, the associated $l$ factor is $3$. Here,
without distinguishing the contribution from each diagram, we take the mean
value $l=2.5$ and associate the uncertainty of $l$ to the error estimation
of $\zeta $.

The above-mentioned constraint, $\delta T_{00}=0$, yields, 

\bigskip 


\begin{figure}[tbp]
\begin{center}
\includegraphics[height=7.5cm]{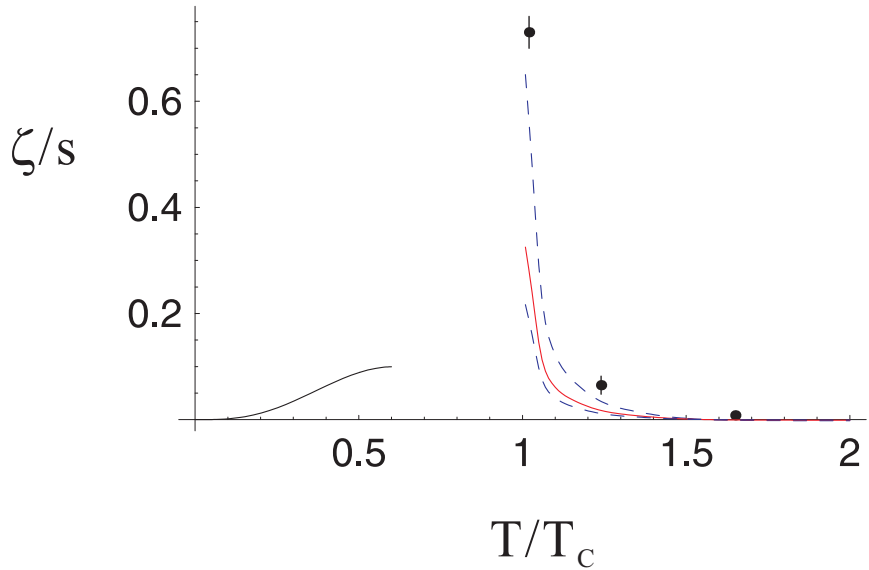}
\end{center}
\caption{(Color online) $\protect\zeta /s$ shown as a function of $T/T_{c}$. 
The 
solid line
below $T_{c}$ is the massless pion gas result ($T_{c}\simeq 200$ MeV and $%
l=2.5$, explained below Eq.(\protect\ref{G}), are used). The error on this
curve is estimated to be $30\%$-$40\%$. The points are the lattice results
for gluon plasma \protect\cite{Meyer:2007dy}. The solid and dashed lines
above $T_{c}$ give the central values and the error band from the QGP sum
rule result of Ref. \protect\cite{Karsch:2007jc}.}
\label{1}
\end{figure}


The dimensionless combination $\zeta /s$ is shown in Fig. 1. The solid line
below $T_{c}$ is the leading order massless pion gas result (we have used $%
l=2.5$, explained below Eq.(\ref{G}), and the lattice result, $T_{c}\simeq
200$ MeV, for 2+1 flavors of improved staggered fermion as an estimation 
\cite{Cheng:2007jq}). The error on this curve is estimated to be $30\%$-$%
40\% $ from $l$ and higher-order corrections. But the monotonic increasing
behavior should be robust. The solid points are the lattice results for
gluon plasma \cite{Meyer:2007dy}. The QGP curves above $T_{c}$ (the solid
line gives the central values and the dashed lines give the estimated
errors) are based on an exact sum rule, a lattice result for the equation of
state, and a spectral function ansatz with massive quarks \cite%
{Karsch:2007jc}. Since the light quark mass dependence in the QGP curve is
expected to be small, Fig. 1 shows that, in the chiral limit, QCD $\zeta /s$
reaches its maximum while $\eta /s$ reaches its minimum around $T_{c}$ as
mentioned above. The same $\zeta /s$ behavior is also seen in
molecular-dynamics simulations of Lennard-Jones model fluids \cite{Meier}.

A recent massive pion gas calculation shows that $\zeta $ has two peaks \cite%
{FernandezFraile:2008vu}, one is near 10 MeV and the other is near $T_{c}$.
They are corresponding to breaking of the conformal symmetry by the pion
mass and the anomaly, respectively. The behavior near the higher temperature
peak is similar to what we have found here for the massless pion case. It is
also similar to the $\zeta $ behavior of \cite{NoronhaHostler:2008ju} near $%
T_{c}$ with Hagedorn states included. The behavior near the lower
temperature peak is similar to earlier results of \cite%
{Prakash:1993bt,Davesne:1995ms}. The massless pion calculation of \cite%
{FernandezFraile:2008vu} also conforms our qualitative behavior of $\zeta $.

In the large $N_{c}$ (the number of colors) limit, 
\begin{equation}
\frac{\zeta }{s}\propto \frac{1}{N_{c}^{2}N_{f}^{2}}\ \ \text{for massless
pion gas,}
\end{equation}%
and%
\begin{equation}
\frac{\zeta }{s}\propto \frac{\alpha _{s}^{2}}{N_{c}^{2}}\propto \frac{1}{%
N_{c}^{4}}\ \ \text{for PQCD,}
\end{equation}%
where we have used the scaling $f_{\pi }\propto \sqrt{N_{c}}$, $g_{\pi
}\propto N_{f}^{2}$, $\alpha _{s}^{2}\propto 1/N_{c}$ and $N_{f}$ is the
number of light quark flavors. Also, for massless pions, 
\begin{equation}
\frac{\zeta }{\eta }\simeq 180\left( \frac{l}{2.5}\right) \left( \frac{1}{3}-%
\frac{{\mathcal{P}}}{\epsilon }\right) \left( \frac{1}{3}-v_{s}^{2}\right) \
.
\end{equation}%
This is similar to $\zeta /\eta \sim 15\left( 1/3-v_{s}^{2}\right) ^{2}$,
which is obtained for a photon gas coupled to hot matter \cite%
{Weinberg:1971mx} and is also parametrically correct for PQCD \cite%
{Arnold:2006fz}. This is because in those cases, $2\rightarrow 2$ scattering
is the dominant process in both $\zeta $ and $\eta $ computations. It is not
the case, however, in $\phi ^{4}$ theory in which $\left(
1/3-v_{s}^{2}\right) ^{-2}\zeta /\eta $ has large $T$ dependence because $%
\zeta $ is dominated by $2\leftrightarrow 4$ scattering while $\eta $ is
dominated by $2\rightarrow 2$ scattering. The scaling is also different from 
$\zeta /\eta \propto \left( 1/3-v_{s}^{2}\right) $ for strongly coupled $%
\mathcal{N}=2^{\ast }$\ gauge theory using AdS/CFT \cite{Benincasa:2005iv}%
.\bigskip

\section{Conclusions}

We have computed the bulk viscosity for a \ gas of massless pions using the
Boltzmann equation with the kinetic theory generalized to incorporate the
trace anomaly. The resulting $\zeta /s$, together with the corresponding
results of gluon plasma \cite{Meyer:2007dy} and quark gluon plasma \cite%
{Kharzeev:2007wb} indicates $\zeta /s$ reaches its maximum near $T_{c}$
while $\eta /s$ reaches its minimum near $T_{c}$. If the $\zeta /s$ behavior
is unchanged for massive pions, then the hadronization of the fire ball in
heavy ion collisions would imply large entropy production \cite%
{Meyer:2007dy,Kharzeev:2007wb} and slow equilibration. It would be
interesting to explore the implications of the possible large bulk viscosity
near a phase transition in cosmology if the phase transition above the TeV
scale is based on some strongly interacting mechanism.

We thank Harvey Meyer and Larry Yaffe for useful discussions. We also thank
Frithjof Karsch, Dmitri Kharzeev, and Kirill Tuchin for providing the table
of their result of Ref. \cite{Kharzeev:2007wb}. This work is supported by
the NSC and NCTS of Taiwan.


\end{document}